\documentclass{PoS}

\title{Kaluza-Klein Dark Matter: Direct Detection vis-a-vis LHC}

\ShortTitle{Kaluza-Klein Dark Matter: Direct Detection vis-a-vis LHC}

\author{\speaker{Sebastian Arrenberg}$^a$, Laura Baudis$^a$, Kyoungchul Kong$^b$, 
	Konstantin T. Matchev$^c$ and Jonghee Yoo$^b$\\
         \llap{$^a$} Physics Institute, University of Z\"urich\\
          \llap{$^b$} Fermi National Accelerator Laboratory\\
          \llap{$^c$} Physics Department, University of Florida\\

        \\
        E-mails: \email{arrenberg@physik.uzh.ch}, \email{laura.baudis@physik.uzh.ch}, \email{kckong@fnal.gov},
        \email{matchev@phys.ufl.edu}, \email{yoo@fnal.gov}}

\abstract{
We explore the phenomenology of Kaluza-Klein (KK) dark matter in 
very general models with universal extra dimensions (UEDs), emphasizing the
complementarity between high-energy colliders and dark matter direct detection experiments.
In models with relatively small mass splittings between the dark matter 
candidate and the rest of the (colored) spectrum, the collider 
sensitivity is diminished, but direct detection rates are enhanced.
UEDs provide a natural framework for such mass degeneracies. 
We consider both 5-dimensional and 6-dimensional non-minimal UED models,
and discuss the detection prospects for various KK dark matter candidates:
the KK photon $\gamma_1$ (5D) the KK $Z$-boson $Z_1$ (5D)
and the spinless KK photon $\gamma_H$ (6D).  
We combine collider limits such as electroweak precision data and expected 
LHC reach, with cosmological constraints from WMAP and the
sensitivity of current or planned direct detection experiments.
Allowing for general mass splittings, 
we show that neither colliders, nor direct detection 
experiments by themselves can explore all of the relevant 
KK dark matter parameter space. Nevertheless, they probe different 
parameter space regions and the combination of the two types of
constraints can be quite powerful. For example, in the case of 
$\gamma_1$ in 5D UEDs the relevant parameter space will be almost completely 
covered by the combined LHC and direct detection sensitivities expected in the near future. 
The work presented here is based on \cite{basic}.}

\FullConference{Identification of dark matter 2008\\
		 August 18-22, 2008\\
		 Stockholm, Sweden}

\begin{document}

\section{Introduction}

In the framework of UEDs \cite{hooper} all standard model (SM) particles are promoted to one or more flat, compactified extra 
dimensions. An infinite number of new particles, called Kaluza-Klein tower, arises  for every SM particle. Since at tree level their masses 
receive a dominant contribution $\sim {\rm TeV}$ arising from the momentum carried along the extra dimension(s) the spectrum is highly 
degenerated and thus radiative corrections are significant. The lightest Kaluza-Klein partner (LKP) is stable and if it is
neutral, it can be a possible dark matter candidate. Considering the mass spectrum 
we did not restrict ourselves to the minimal UED (MUED) framework where vanishing boundary interactions at the cut-off scale are assumed, but rather 
allowed for a more general scenario taking the mass splitting 

\begin{displaymath}
\Delta_{q_1} = \frac{m_{q_1}-m_{LKP}}{ m_{LKP}} 
\end{displaymath}
between the LKP and the level one KK quarks as a free parameter.

\section{Relic density calculations}

Since a high degree of mass degeneracy occurs quite naturally in models with UEDs it is of particular 
importance to take coannihilations into account when computing the relic density of the LKP candidate \cite{relic}.
The procedure is straightforward. After all heavier particles have decayed into it, the number
density of the lightest species obeys a simple Boltzmann equation, where essentially the cross section 
has to be replaced by an effective cross section which depends on all annihilation cross sections and the mass 
splittings between the lightest particle and all other particles considered.

\begin{figure}[b]
\centering
\includegraphics[width=0.48\textwidth]{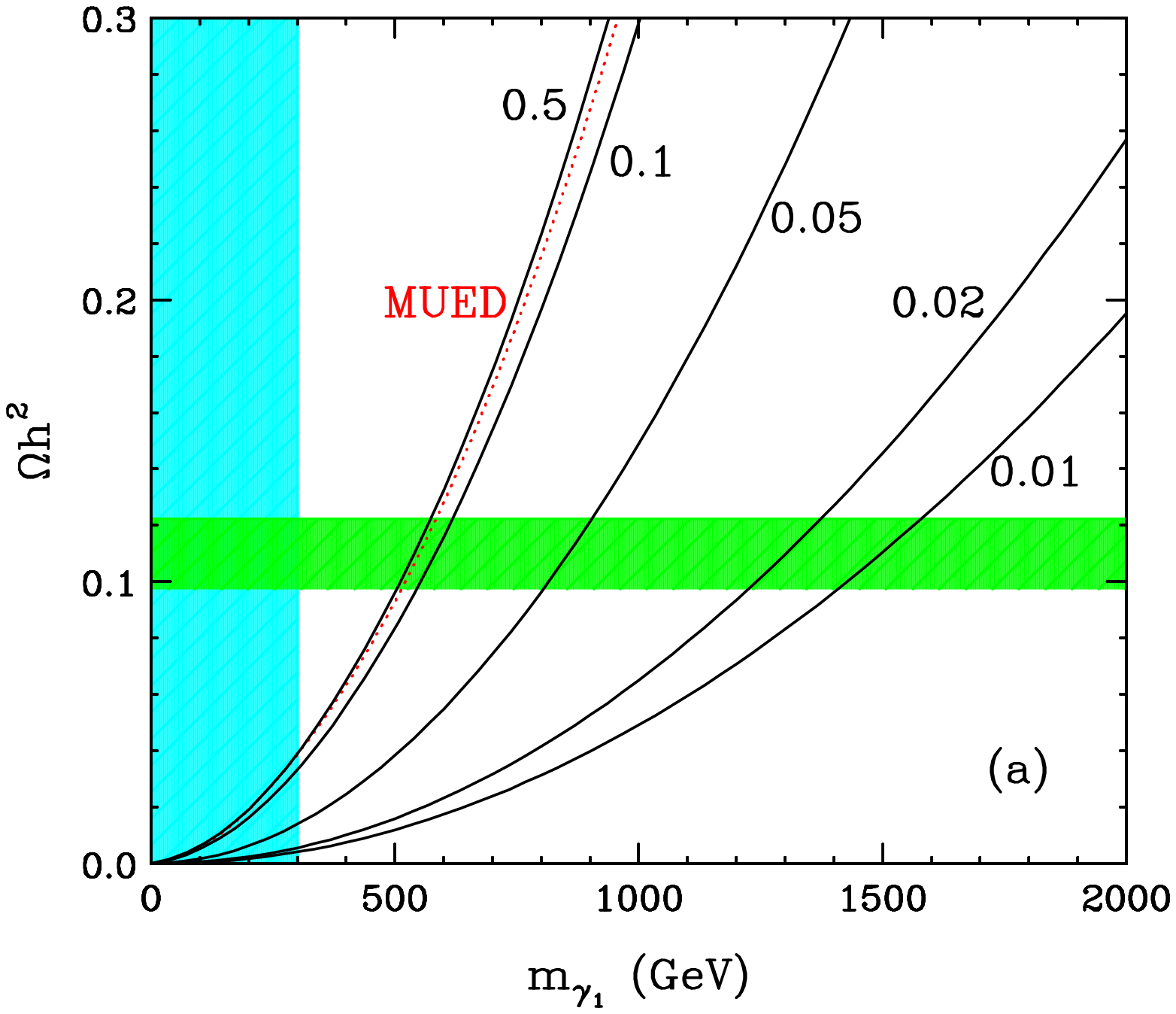}
\hspace{0.12cm}
\includegraphics[width=0.48\textwidth]{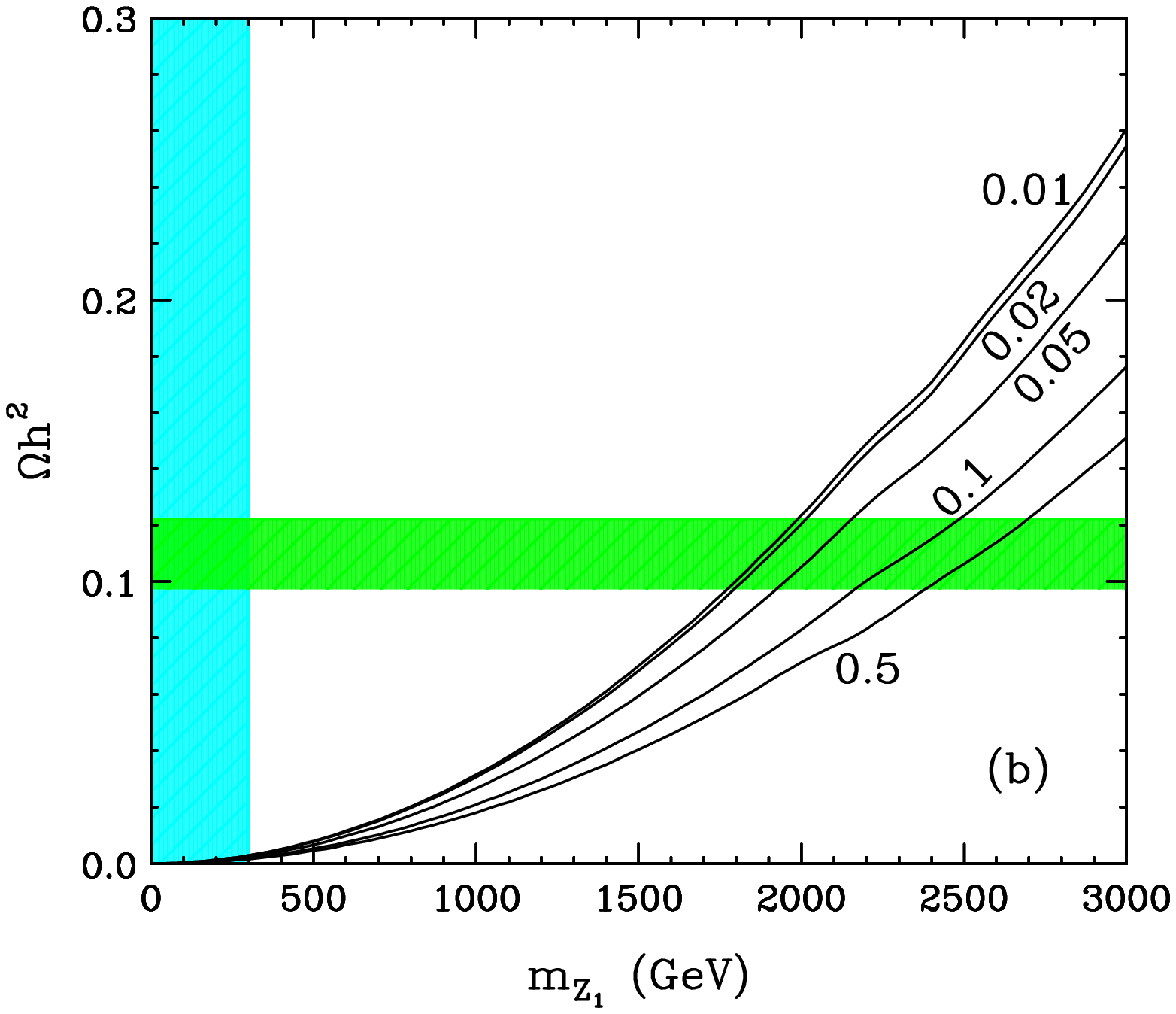}
\caption{\sl Relic density of the LKP and constraints from electroweak precision data and WMAP.}
\label{fig:Omegah2_B1_Z1}
\end{figure}

We computed the relic densities for $\gamma_1$ \cite{relic} and $Z_1$ in 5D UED taking coannihilations with all other level 
one KK particles into account. In the MUED framework the $\gamma_1$ is the LKP. The result
of the computation is shown as a dotted, red line in Fig.~\ref{fig:Omegah2_B1_Z1}(a). In a second approach a certain mass 
splitting $\Delta$ between the LKP and the KK quarks was assumed fixing the rest of the spectrum at their MUED masses. Each line in
the plot is labeled with the corresponding value of $\Delta$. Coannihilations thus decrease the predictions for the relic
density. In the case of the $Z_1$ shown in Fig.~\ref{fig:Omegah2_B1_Z1}(b) the quark masses were fixed in the same way 
as before. $Z_1$ and $W_1^\pm$ are assumed to be degenerate while the gluon is heavier than $Z_1$ by 10\% and 
all other KK particles are heavier by 20\%. In this case the effect of coannihilations is inverted showing that its
sign cannot  easily be predicted. In both plots we also show constraints from electroweak precision data 
\cite{electroweak1, electroweak2} as a cyan vertical band and the preferred 2$\sigma$-WMAP region \cite{wmap} as a green horizontal
band. For example, a mass of 500 GeV for the $\gamma_1$ LKP in 5D MUED is a good 
benchmark.\footnote{Note that the relic density including coannihilations of $\gamma_H$ in 6D UED has not been investigated yet.}

\section{Direct LKP Detection - Predictions and Limits}

In order to explore constraints from direct detection experiments it is necessary to investigate the elastic scattering between 
the dark matter particle and target nuclei. For all three LKP candidates considered here, $\gamma_1$ and $Z_1$ in 5D UED
and $\gamma_H$ in 6D UED, there are two Feynman diagrams arising from KK quark exchange and one from Higgs boson
exchange, contributing to the scattering cross section at tree level \cite{basic}. Here we only consider spin-independent
scattering.

\begin{figure}[b!]
\centering
\includegraphics[width=0.57\textwidth]{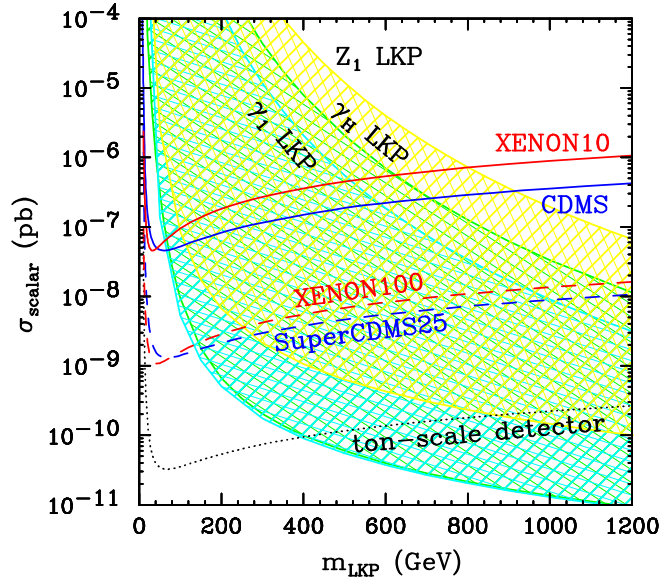}
\caption{\sl Theoretical predictions for spin-independent LKP-nucleon cross sections and sensitivities from current and
planned future direct detection experiments.}
\label{fig:SI_CrossSection_Neutron}
\end{figure}

The theoretical predictions are shown in Fig.~\ref{fig:SI_CrossSection_Neutron}. We fixed the Higgs 
mass at 120 GeV and assumed $\Delta$ to be between 1\% which is the upper boundary of the respective shaded area and 50\% 
which is the lower boundary. The plot also contains cross section limits from CDMS \cite{cdms} and XENON10 \cite{xenon} as well as expected sensitivities from
future experiments. Current experiments already exclude small mass splittings while future experiments should cover most of the 
relevant parameter space.

\section{Limits on Kaluza-Klein Dark Matter}

\begin{figure}[b!]
\centering
\includegraphics[width=0.47\textwidth]{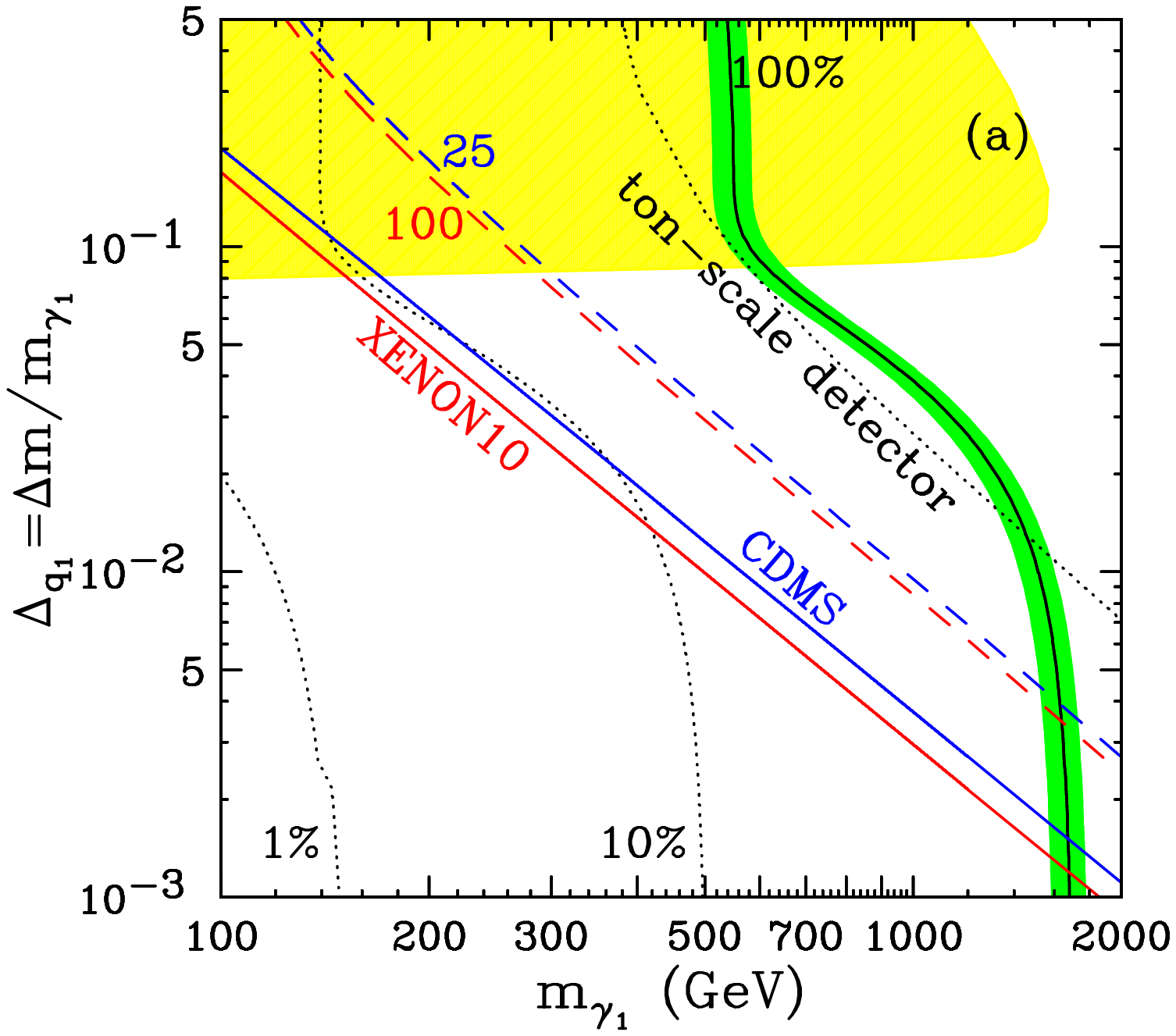}
\hspace{0.17cm}
\includegraphics[width=0.47\textwidth]{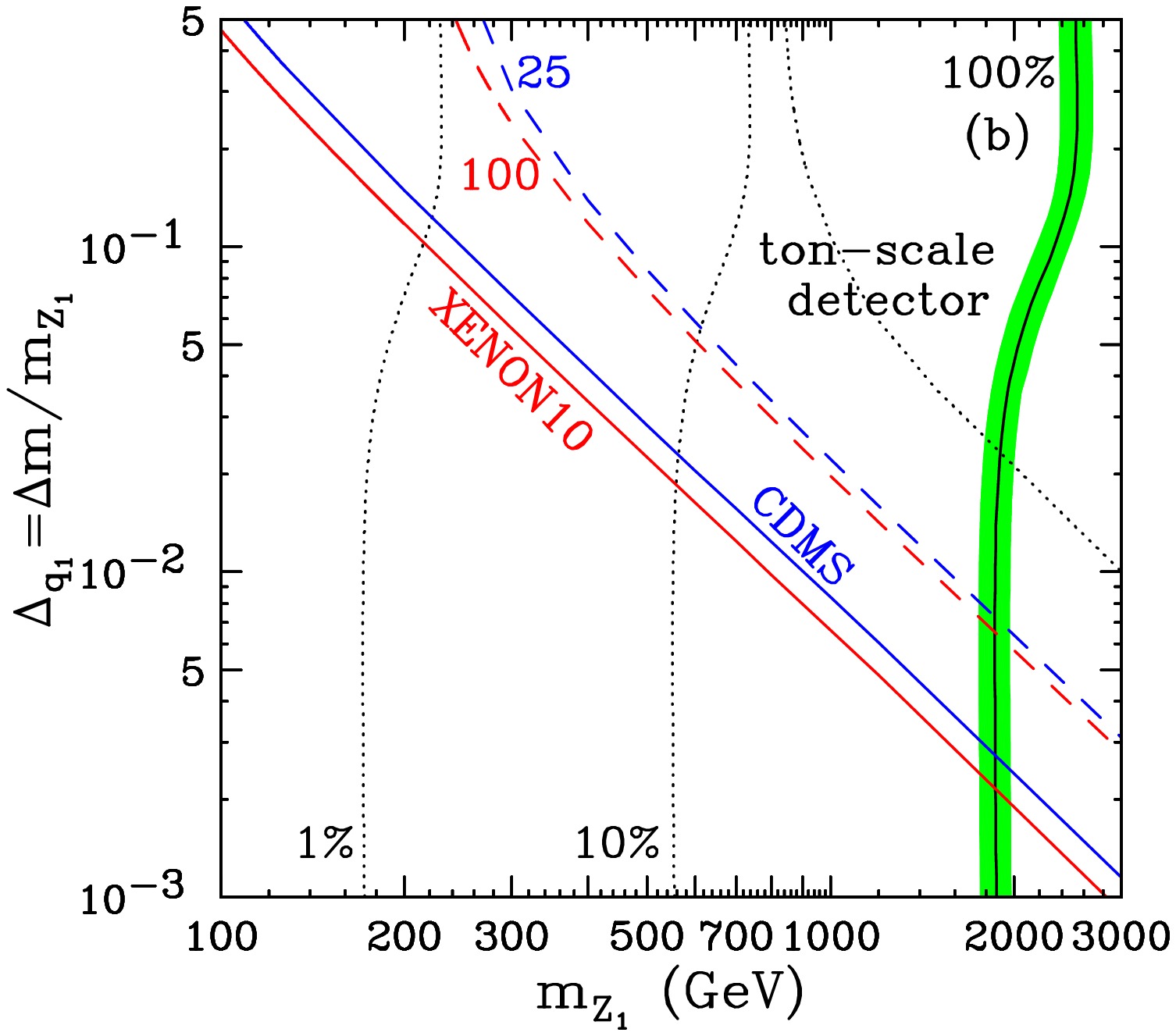}
\caption{\sl Limits from direct detection experiments, LHC studies and WMAP  in the $\Delta$ vs. $m_{LKP}$ plane.
}
\label{fig:SI_Delta_Neutron_B_Z}
\end{figure}

In addition to calculating the limits on the spin-independent cross sections we also performed a more detailed analysis of the LKP specific parameter space.

For example, one can translate the information from Fig.~\ref{fig:SI_CrossSection_Neutron} into the $\Delta$ vs. WIMP mass plane when fixing the
Higgs mass. This is shown in Fig.~\ref{fig:SI_Delta_Neutron_B_Z} for $\gamma_1$ and $Z_1$ in 5D UED. All mass splittings below the
respective limit curve are excluded. As expected direct detection experiments are particularly sensitive to small $\Delta$'s. Additionally
both plots contain relic density constraints providing upper limits
on the LKP mass. The black solid line accounts for all the dark matter in the universe while the two black dotted lines show the bounds assuming
that the LKP would contribute only 1\% or 10\% to the total amount of dark matter. The green shaded region represents the preferred 
2$\sigma$-WMAP region \cite{wmap}.
For the case of the $\gamma_1$ a collider study of the $4\ell + \rlap{\,\,/}E_T$ channel yielded the result that with a luminosity of 100~fb$^{-1}$
the yellow shaded region should be covered by the LHC \cite{lhc}. The LHC is thus more sensitive to large mass splittings. All three probes are highly 
complementary  and in the case of the $\gamma_1$ the entire parameter space could be covered by ton-scale direct detection experiments.

\begin{figure}[t]
\centering
\includegraphics[width=0.39\textwidth]{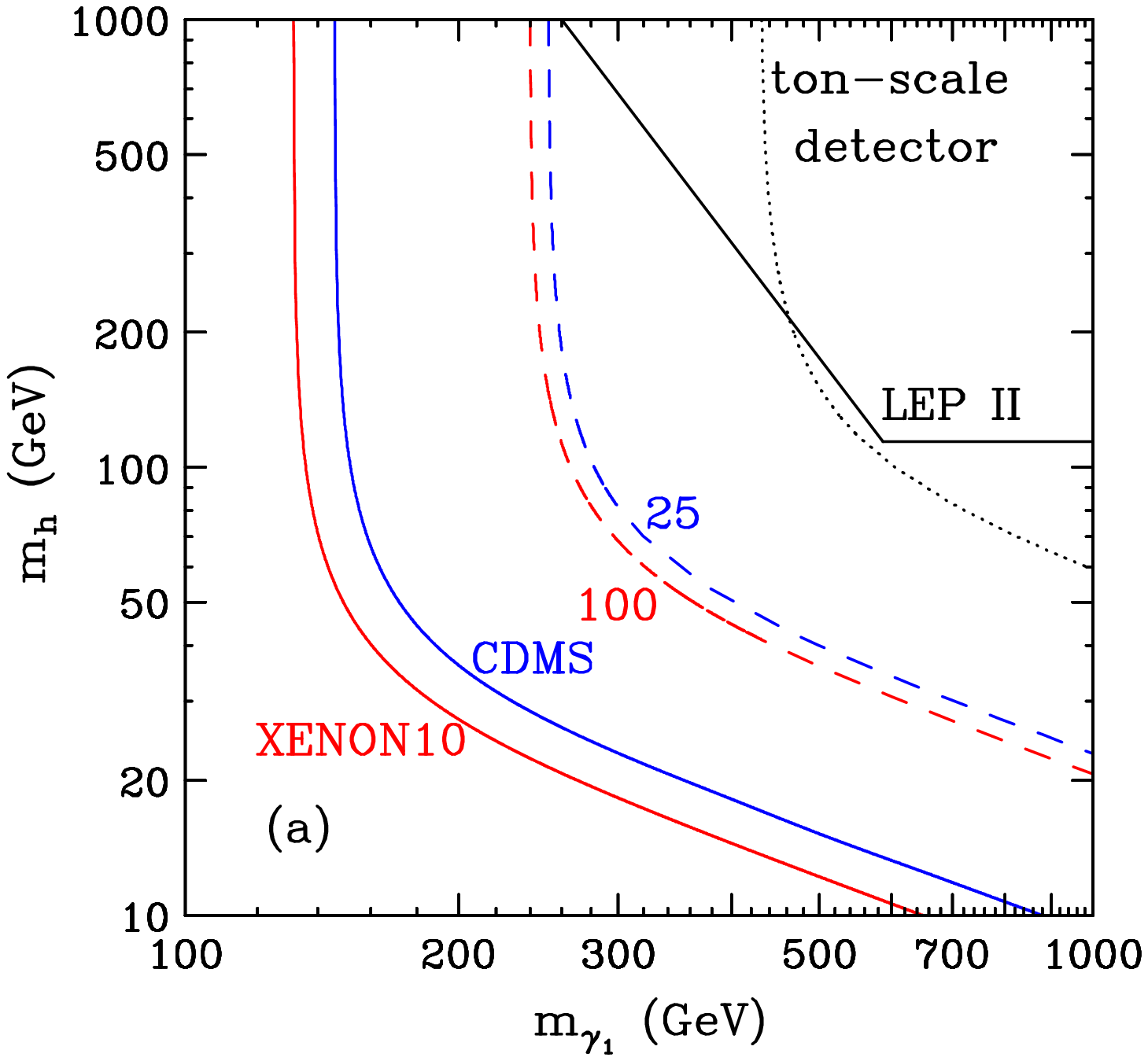}
\hspace{0.5cm}
\includegraphics[width=0.39\textwidth]{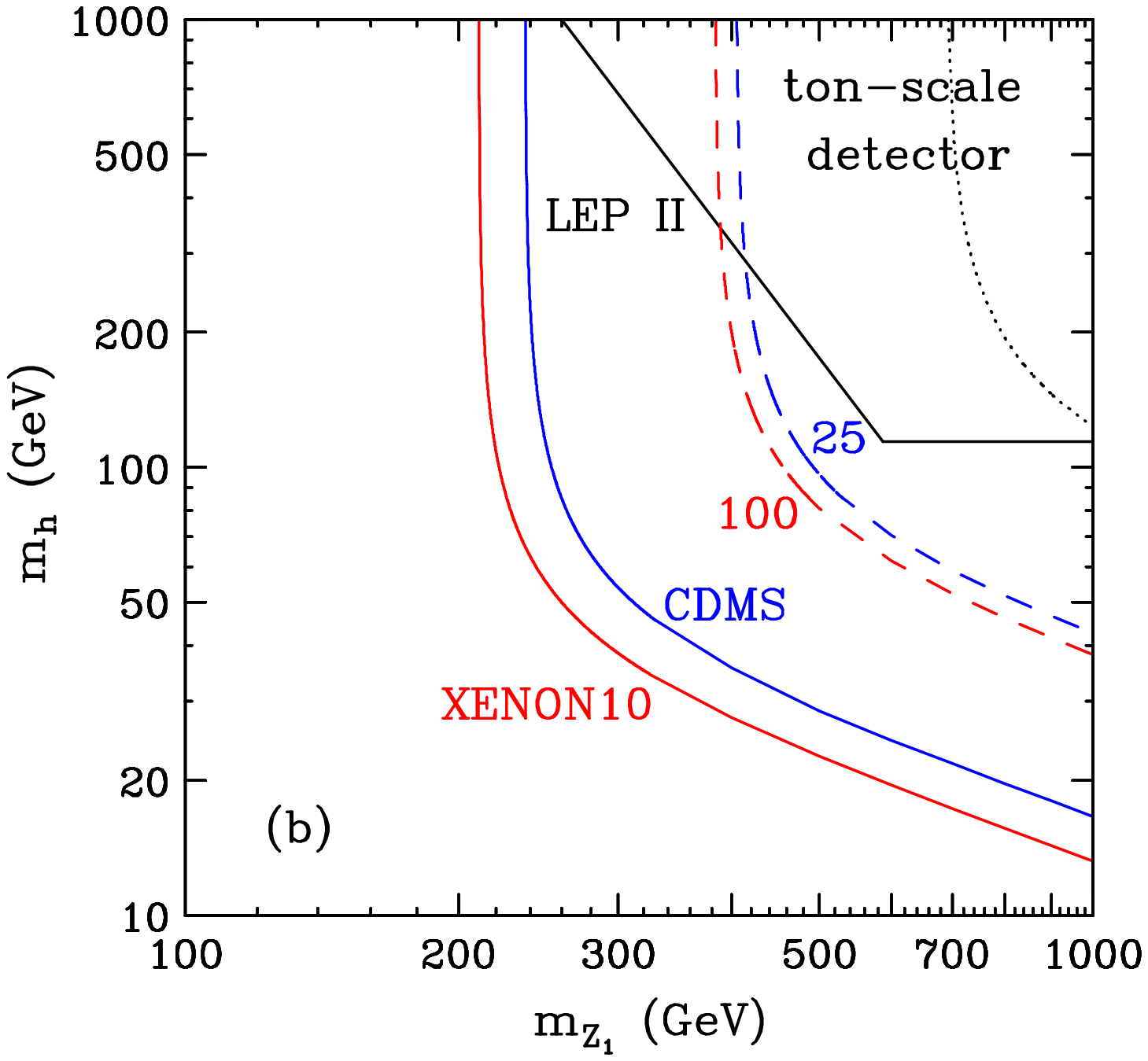}
\caption{\sl Limits from direct detection experiments and collider studies in the $m_h$ vs. $m_{LKP}$ plane.}
  \label{fig:SI_HiggsMass_Neutron_B_Z}
\end{figure}

On the other hand, one can also interchange the role of the mass splitting and the Higgs mass. In Fig.~\ref{fig:SI_HiggsMass_Neutron_B_Z}
we show limits on the Higgs mass fixing $\Delta$ at 10\%. The excluded regions are below and to the left of each limit curve. The
asymptotic behaviour is related to the decoupling of the Higgs exchange for large Higgs masses. In each plot the horizontal black line 
represents the current Higgs mass bound of 114 GeV while the diagonal line shows the limit from oblique  corrections 
\cite{electroweak2}.\footnote{Note that these corrections were computed considering the $\gamma_1$ to be the LKP in MUED. Thus the corresponding bound in 
Fig.~\ref{fig:SI_HiggsMass_Neutron_B_Z}(b) has been included only for illustrative reasons.}
From the observation that the Higgs mass dependence in direct detection experiments only shows up in already excluded parts of the parameter
space it can be concluded that it plays a secondary role interpreting those experiments. Even future direct detection experiments
will only probe a small part of the relevant parameter space, however the LHC will be more sensitive with respect to the Higgs mass.

\noindent
For a more detailed analysis, including the parameter space of $\gamma_H$ LKP in 6D UED and spin-dependent interactions, we
refer to \cite{basic}.

\section{Conclusion}

We performed a comprehensive phenomenological analysis of Kaluza-Klein dark matter, including constraints from direct detection experiments,
collider studies and cosmology considering 5D and 6D UED. It was shown that all three approaches are highly complementary and combining them
substantially diminishes the relevant parameter space. Direct detection experiments restrict small values of $\Delta$ whereas colliders are sensitive
to large mass splittings. Cosmology on the other hand rules out large LKP masses and as shown here these two parameters ($\Delta$ and $m_{LKP}$)
are the relevant quantities in analyzing UEDs. Moreover the importance of coannihilations for relic density computations should be emphasized. 

Important investigations for the future in the context of UEDs would be further LHC and more detailed relic density studies for certain LKP candidates
such as the $\gamma_H$ LKP in 6D UED.


\begin{thebibliography}{99}
	\bibitem{basic} 
 	S.~Arrenberg, L.~Baudis, K.~Kong, K.T.~Matchev and J.~Yoo, Phys. Rev. D {\bf 78}, 056002 (2008) [{\tt hep-ph/0805.4210}]

	\bibitem{hooper} 
 	D.~Hooper and S.~Profumo, Phys.\ Rept.\  {\bf 453}, 29 (2007) [{\tt hep-ph/0701197}]

 	\bibitem{relic} 
 	K.~Kong and K.T.~Matchev, JHEP  {\bf 0601}, 038 (2006) [{\tt hep-ph/0509119}]
	
	\bibitem{electroweak1} 
 	T.~Appelquist and H.~U.~Yee, Phys. Rev. D {\bf 67}, 055002 (2003) [{\tt hep-ph/0211023}]

	\bibitem{electroweak2} 
 	I.~Gogoladze and C.~Macesanu, Phys. Rev. D {\bf 74}, 093012 (2006) [{\tt hep-ph/0605207}]

	\bibitem{wmap} 
 	J.~Dunkley {\it et al.}  (WMAP Collaboration),  [{\tt astro-ph/0803.0586}]

	\bibitem{cdms} 
	Z.~Ahmed {\it et al.}  (CDMS Collaboration),  [{\tt astro-ph/0802.3530}]

	\bibitem{xenon} 
 	J.~Angle {\it et al.}  (XENON10 Collaboration), Phys. Rev. Lett. {\bf 100}, 021303 (2008) [{\tt astro-ph/0706.0039}]

  	\bibitem{lhc} 
 	H.~C.~Cheng, K.T.~Matchev and M.~Schmaltz, Phys. Rev. D {\bf 66}, 056006 (2002) [{\tt hep-ph/0205314}]
	
  \end{thebibliography}
\end{document}